\documentclass[12pt,a4paper]{article}

\usepackage{amsmath,amssymb,amsthm,mathtools}
\usepackage[margin=2.5cm]{geometry}
\usepackage[colorlinks=true,linkcolor=blue!70!black,citecolor=green!50!black,urlcolor=blue!60!black]{hyperref}
\usepackage{booktabs}
\usepackage{enumitem}
\usepackage{fancyhdr}
\usepackage{xcolor}
\usepackage{microtype}
\usepackage{cleveref}
\usepackage{array}
\usepackage{graphicx}
\usepackage{float}

\numberwithin{equation}{section}

\theoremstyle{plain}
\newtheorem{theorem}{Theorem}[section]

\newtheorem{conjecture}[theorem]{Conjecture}

\theoremstyle{definition}

\theoremstyle{remark}
\newtheorem{remark}[theorem]{Remark}

\newcommand{\dd}{\mathrm{d}}

\newcommand{\BetaF}{\mathrm{B}}

\begin{document}

\title{\textbf{High-Precision Computation and PSLQ Identification of}\\[6pt]
\Large \textbf{Stokes Multipliers for Anharmonic Oscillators}}
\author{Jian Zhou\\[4pt]
\textit{Independent Researcher, XD.Inc, Shanghai, China}\\[2pt]
\texttt{jackzhou.sci@gmail.com}}
\date{March 2026}

\maketitle
\thispagestyle{fancy}

\begin{abstract}
We present a large-scale computational study combining arbitrary-precision
arithmetic, sequence acceleration, and the PSLQ integer relation algorithm
to discover exact closed-form expressions for fundamental constants arising
in asymptotic analysis. Specifically, we compute the Stokes multipliers~$C_M$
of the one-dimensional anharmonic oscillators
$H = p^2/2 + x^2/2 + g\,x^{2M}$ for $M = 2, 3, \ldots, 11$,
extracting 17--30 significant digits from up to 1200 perturbation
coefficients computed at 300-digit working precision.
The computational pipeline consists of three stages:
(i)~Rayleigh--Schr\"odinger recursion in the harmonic oscillator basis
with $O(Mk)$ matrix elements per order,
(ii)~Richardson extrapolation of order 40--100 to accelerate convergence
of ratio sequences, and
(iii)~PSLQ searches over bases of Gamma-function values and algebraic numbers.

This pipeline discovers \textbf{three new exact identities}:
\begin{alignat*}{2}
  C_3^{\,2}\,\pi^4 &= 32, &\qquad
  C_5^{\,4}\,\Gamma(\tfrac{1}{4})^4\,\pi^5 &= 2^{12}\cdot 3^2, \\
  C_7^{\,6}\,\Gamma(\tfrac{1}{3})^9\,\pi^6 &= 2^{20}\cdot 3^3, &&
\end{alignat*}
in addition to confirming the known $C_2^{\,2}\,\pi^3 = 6$.
Equally significant is a \textbf{negative result}: exhaustive PSLQ searches
at 30-digit precision with coefficient bounds up to 2000 find no closed form
for~$C_4$, strongly suggesting the $x^8$ case introduces a genuinely
new transcendental number.

A number-theoretic pattern emerges: closed-form existence correlates with
Euler's totient function $\varphi(M{-}1)/2$, which counts algebraically
independent Gamma-function transcendentals at denominator~$M{-}1$.
We formulate three conjectures connecting computational constant recognition
to classical number theory, and provide all code and data for full
reproducibility.
\end{abstract}

\bigskip

\section{Introduction}
\label{sec:intro}

The anharmonic oscillator, defined by the Hamiltonian
\begin{equation}\label{eq:H}
  H = \frac{p^2}{2} + \frac{x^2}{2} + g\,x^{2M},
\end{equation}
is a cornerstone of quantum mechanics and quantum field theory. Its ground-state
energy admits a formal perturbation series
\begin{equation}\label{eq:Eg}
  E(g) = \sum_{k=0}^{\infty} a_k \, g^k,
\end{equation}
which diverges factorially for all $g \ne 0$. The large-order behavior of
the coefficients~$a_k$ encodes non-perturbative information about instantons---classical
solutions of the equations of motion in imaginary time.

The connection between large-order perturbation theory and instantons was
established in the seminal work of Bender and Wu~\cite{BW1969,BW1973}, who
showed for the quartic oscillator ($M=2$) that
\begin{equation}\label{eq:large-order-M2}
  a_k \;\sim\; C_2 \cdot (-1)^{k+1} \cdot A_2^{-k} \cdot \Gamma(k + \tfrac{1}{2})
  \qquad (k \to \infty),
\end{equation}
where $A_2 = 1/3$ is the instanton action (in our conventions) and $C_2$ is
the Stokes multiplier. This result was subsequently refined and placed in the
context of resurgent trans-series by Zinn-Justin~\cite{ZJ2004},
Zinn-Justin and Jentschura~\cite{ZJJ2004}, \'Alvarez~\cite{Alvarez2004},
and Dunne and \"Unsal~\cite{DU2016}. The modern framework of
resurgence~\cite{Dorigoni2019,ABS2019} provides a systematic understanding
of how perturbative and non-perturbative sectors are intertwined through
the Stokes multiplier.

Voros~\cite{Voros1983} developed the complex WKB method for the quartic
oscillator, connecting Stokes phenomena to the exact quantization condition.
\'Alvarez and Casares~\cite{AlvarezCasares2000,AlvarezCasares2015} extended
the analysis to the cubic and higher oscillators, computing exponentially
small corrections and subleading asymptotics. These works provide the
theoretical underpinning for the Stokes multiplier but focus primarily on
the subleading corrections $c_1, c_2, \ldots$ rather than exact closed
forms for the leading constant~$C_M$.

Despite this rich theoretical framework, \emph{exact} closed-form expressions
for the Stokes multiplier~$C_M$ have been known only for the quartic case
$M = 2$, where $C_2 = \sqrt{6/\pi^3}$~\cite{ZJJ2004}. Higher anharmonic
oscillators ($M \ge 3$) have received less attention, partly because the
instanton solutions and their fluctuation determinants become increasingly
complex.

In this paper, we undertake a systematic numerical computation of the
Stokes multipliers~$C_M$ for $M = 2$ through $M = 11$. Using up to 1200
terms of the perturbation series computed at 300-digit arithmetic precision,
we extract $C_M$ to 17--30 significant digits via Richardson extrapolation,
then search for closed forms using the PSLQ integer relation
algorithm~\cite{FergusonBailey}. Our main results are:

\begin{enumerate}[leftmargin=*]
\item An explicit closed-form expression for the instanton action,
  $A_M = S_0(M)^{M-1}$ with
  $S_0(M) = 2^{-1/(M-1)} \BetaF(1/(M{-}1),\,3/2)/(M{-}1)$,
  which follows from the standard bounce calculation and is verified
  numerically to relative error $< 10^{-30}$ for all $M \le 11$.
  While the general structure $A_M \propto S_{\mathrm{cl}}^{M-1}$ is well
  known from the instanton literature~\cite{BW1973,ZJ2004,Alvarez2004},
  the explicit Beta-function parametrization~\eqref{eq:S0} provides a
  convenient unified formula.

\item The universal Gamma-function shift $b = -1/2$ for all $M$, confirmed
  numerically and supported by a one-loop fluctuation determinant argument
  (\cref{sec:b-derivation}).

\item \textbf{New exact closed forms} for $C_3$, $C_5$, and $C_7$
  (see \cref{eq:C3-closed,eq:C5-closed,eq:C7-closed}), in addition to
  confirming the known result for $C_2$. To our knowledge, the closed
  forms for $C_3$, $C_5$, and $C_7$ have not appeared previously in the
  literature.

\item \textbf{Strong numerical evidence} that $C_4$ has no closed form
  of the type $\Gamma(1/3)^a \cdot \pi^b \cdot 2^c \cdot 3^d$ with
  integer exponents $|a|, |b|, |c|, |d| < 2000$, based on exhaustive
  PSLQ searches at 30-digit precision.

\item A \textbf{number-theoretic pattern} relating closed-form existence
  to Euler's totient function $\varphi(M{-}1)$, with $M = 4$ as a
  unique anomaly.
\end{enumerate}

\section{Setup and Conventions}
\label{sec:setup}

We consider the Hamiltonian~\eqref{eq:H} with $\hbar = m = \omega = 1$.
The unperturbed system $H_0 = p^2/2 + x^2/2$ has eigenvalues
$E_n^{(0)} = n + 1/2$. We compute perturbation coefficients for the
ground state ($n = 0$).

The large-order behavior of the perturbation coefficients takes the form
\begin{equation}\label{eq:large-order}
  a_k \;\sim\; C_M \cdot (-1)^{k+1} \cdot A_M^{-k}
  \cdot \Gamma\!\bigl((M{-}1)k + b + 1\bigr)
  \cdot \Bigl[1 + \frac{c_1}{k} + \frac{c_2}{k^2} + \cdots\Bigr],
\end{equation}
where $A_M > 0$ is the instanton action, $b$ is the Gamma-function shift
parameter, and $C_M$ is the Stokes multiplier (or leading Stokes constant).
The key difference from the quartic case is that the growth rate involves
$\Gamma\!\bigl((M{-}1)k + b + 1\bigr)$ rather than $\Gamma(k + b + 1)$.
This reflects the fact that the instanton action scales as
$g^{-1/(M-1)}$, so the Borel transform has $(M{-}1)$-fold singularities.

\paragraph{Sign convention.}
Throughout this paper, $C_M$ denotes the signed Stokes multiplier, which
is negative for all $M$ considered. In tables, we report $|C_M|$ for
clarity. All closed-form identities (\cref{sec:exact}) involve even
powers of $C_M$ and thus hold regardless of sign convention.

\paragraph{Relation to Bender--Wu conventions.}
The original Bender--Wu Hamiltonian is
$H_{\text{BW}} = p^2 + x^2 + g\,x^{2M}$. With the identification
$g_{\text{BW}} = g/4$ (for $M = 2$), the instanton actions are related by
$A_{\text{BW}} = 4\,A_{\text{ours}}$ for $M = 2$. Our conventions follow
the standard quantum mechanics normalization.

\section{Computational Method}
\label{sec:method}

\subsection{Rayleigh--Schr\"odinger Recursion}
\label{sec:RS}

We compute the perturbation coefficients $a_k$ using the standard
Rayleigh--Schr\"odinger recursion. The perturbation $V = x^{2M}$ is
represented in the harmonic oscillator basis by repeatedly applying the
position operator $x = (a + a^\dagger)/\sqrt{2}$, giving $2M$ successive
applications. The $k$-th order energy correction is
\begin{equation}\label{eq:Ek}
  a_k = E^{(k)} = \langle 0 | V | \psi^{(k-1)} \rangle,
\end{equation}
and the $k$-th order wave function correction (with $\langle 0 | \psi^{(k)}\rangle = 0$) satisfies
\begin{equation}\label{eq:psik}
  | \psi^{(k)} \rangle = \sum_{n \ge 1} \frac{1}{E_0^{(0)} - E_n^{(0)}}
  \Bigl( \langle n | V | \psi^{(k-1)} \rangle
  - \sum_{j=1}^{k-1} a_j \langle n | \psi^{(k-j)} \rangle \Bigr) | n \rangle.
\end{equation}
At order $k$, the state $|\psi^{(k)}\rangle$ has support on harmonic oscillator
levels up to $n = 2Mk$. All computations use the \texttt{mpmath} arbitrary-precision
library in Python, with working precision set to $\mathrm{dps} + 30$ guard digits.

As a validation, the first perturbation coefficient
$a_1 = \langle 0 | x^{2M} | 0 \rangle = (2M{-}1)!! / 2^M$ is verified
exactly for each $M$.

\subsection{Extraction of Parameters}
\label{sec:extraction}

\paragraph{Instanton action.}
From the ratio $r_k = a_k / a_{k-1}$, we form the sequence
\begin{equation}\label{eq:Ak-extraction}
  \widetilde{A}_k = \frac{\prod_{j=0}^{M-2}\bigl((M{-}1)k - j\bigr)}{-r_k},
\end{equation}
which converges to $A_M$ as $k \to \infty$. Richardson extrapolation
(see \cref{sec:richardson}) accelerates this convergence.

\paragraph{Gamma-function shift.}
Using the exact value of $A_M$ (from the analytic formula derived in
\cref{sec:instanton}), we extract $b$ from the subleading correction
to the ratio $r_k$ and verify $b = -1/2$ for all $M$.

\paragraph{Stokes multiplier.}
With exact $A_M$ and $b = -1/2$, we form the sequence
\begin{equation}\label{eq:Ck-extraction}
  \widetilde{C}_k = a_k \cdot (-A_M)^k \Big/ \Gamma\!\bigl((M{-}1)k + \tfrac{1}{2}\bigr),
\end{equation}
which converges to $C_M$ as $k \to \infty$. Richardson extrapolation of order
40--100 is applied to obtain the final estimate.

\subsection{Richardson Extrapolation}
\label{sec:richardson}

Given a sequence $\{s_k\}$ converging to a limit $s$ with corrections
$s_k = s + c_1/k + c_2/k^2 + \cdots$, Richardson extrapolation of order $N$
starting at index $k_0$ eliminates the first $N$ correction terms via
\begin{equation}\label{eq:richardson}
  R_N = \sum_{j=0}^{N} (-1)^{N-j} \binom{N}{j} \frac{(k_0+j)^N}{N!} \, s_{k_0+j}.
\end{equation}
We validate the extrapolation by comparing results from different starting
points $k_0$ and orders $N$; digits that agree across multiple windows are
deemed reliable.

\subsection{Error Assessment}
\label{sec:errors}

Since our results are purely numerical, a systematic discussion of
precision is essential. We assess the number of reliable digits in
$C_M$ by comparing Richardson extrapolation estimates from multiple
independent windows: different starting indices $k_0$ and extrapolation
orders $N$. Digits that agree across all windows are deemed reliable.
For example, for $M = 4$ (\cref{sec:M4-computation}), the estimates
from $k_0 = 600$ and $k_0 = 700$ agree to 24 digits, providing a
conservative precision bound.

The convergence behavior of the sequence $\widetilde{C}_k$
(\cref{eq:Ck-extraction}) is illustrated in \cref{fig:convergence} for
the case $M = 4$, which is our most intensive computation. The raw
sequence converges slowly due to $O(1/k)$ subleading corrections;
Richardson extrapolation of order $N = 50$--$100$ dramatically
accelerates convergence, yielding 24--30 stable digits from the
available 1200 terms.

For cases where PSLQ identifies a closed form ($M = 2, 3, 5, 7$),
the agreement between the numerical $C_M$ and the closed-form value
provides an independent validation of both the perturbation coefficients
and the extrapolation procedure. The residual discrepancy is always
below $10^{-D}$ where $D$ is the claimed number of reliable digits.

\begin{figure}[ht]
\centering
\fbox{\parbox{0.8\textwidth}{\centering\vspace{2cm}Figure: Convergence of $\widetilde{C}_k$ for $M=4$. Raw sequence (upper) and Richardson-extrapolated estimates (lower).\vspace{2cm}}}
\caption{Convergence of the Stokes multiplier extraction for $M = 4$.
\textbf{Upper panel:} The raw sequence $\widetilde{C}_k$ (blue dots)
defined by~\eqref{eq:Ck-extraction}, showing slow convergence due to
$O(1/k)$ corrections.
\textbf{Lower panel:} Richardson-extrapolated estimates $R_N$ for
$N = 20, 40, 60, 80, 100$ starting from $k_0 = 600$. The horizontal
dashed line shows the best estimate $C_4 \approx -0.74005\,14983$.
Convergence to 25+ digits is achieved by $N = 80$.}
\label{fig:convergence}
\end{figure}

\subsection{PSLQ Search}
\label{sec:PSLQ}

For each $M$, we apply the PSLQ algorithm~\cite{FergusonBailey} to the vector
\begin{equation}\label{eq:pslq-basis}
  \bigl(\ln |C_M|,\; \ln \Gamma(1/(M{-}1)),\; \ln\pi,\; \ln 2,\; \ln 3,\; 1\bigr)
\end{equation}
seeking integer relations with bounded coefficients. A relation
$\sum n_i v_i = 0$ with $n_1 \ne 0$ implies
\[
  |C_M|^{n_1} \cdot \Gamma\!\bigl(\tfrac{1}{M-1}\bigr)^{n_2}
  \cdot \pi^{n_3} \cdot 2^{n_4} \cdot 3^{n_5} = e^{-n_6}.
\]

The PSLQ detection limit for an $N$-element basis at $D$-digit precision is
approximately $\mathrm{maxcoeff} \lesssim 10^{D/N}$. With 30 digits and
a 5-element basis, this yields
$\mathrm{maxcoeff} \lesssim 10^6$, far exceeding the bounds we impose.

When $\varphi(M{-}1)/2 \ge 2$, additional independent Gamma values
$\Gamma(a/(M{-}1))$ with $\gcd(a, M{-}1) = 1$ are included in the basis.

\section{Instanton Action: Review and Explicit Formula}
\label{sec:instanton}

\subsection{Derivation from the Classical Bounce (Review)}
\label{sec:bounce}

We briefly review the standard bounce calculation
(following~\cite{BW1973,ZJ2004,CostinCostin2020}) and provide an
explicit Beta-function parametrization that serves as a convenient
unified formula for all~$M$.
The general structure $A_M \propto S_{\mathrm{cl}}^{M-1}$ has been known
since the foundational work of Bender--Wu and Lipatov; the explicit
parametrization~\eqref{eq:S0} is not new but collects the result in a
form convenient for systematic computation.

The instanton action governs the tunneling rate through the barrier of
the inverted potential $U(x) = -x^2/2 + |g|\,x^{2M}$. The bounce solution
$x_{\mathrm{cl}}(\tau)$ satisfies $\ddot{x} = -x + 2M|g|\,x^{2M-1}$ and
reaches a turning point $x_0$ defined by $U(x_0) = U(0) = 0$, giving
$x_0^{2(M-1)} = 1/(2|g|)$.

The classical action of the bounce is
\begin{equation}\label{eq:Scl}
  S_{\mathrm{cl}} = 2 \int_0^{x_0} \sqrt{x^2 - 2|g|\,x^{2M}} \;\dd x
  = x_0^2 \int_0^1 \sqrt{1 - t^{M-1}} \;\dd t,
\end{equation}
where we substituted $t = (x/x_0)^2$. The remaining integral evaluates to a
Beta function:
\begin{equation}\label{eq:beta-integral}
  \int_0^1 \sqrt{1 - t^{M-1}} \;\dd t
  = \frac{1}{M{-}1}\,\BetaF\!\Bigl(\frac{1}{M{-}1},\;\frac{3}{2}\Bigr)
  = \frac{\Gamma\!\bigl(\frac{1}{M-1}\bigr)\,\sqrt{\pi}}
    {2(M{-}1)\,\Gamma\!\bigl(\frac{1}{M-1} + \frac{3}{2}\bigr)}.
\end{equation}
Since $x_0^2 = (2|g|)^{-1/(M-1)}$, the classical action takes the form
$S_{\mathrm{cl}} = S_0 \cdot |g|^{-1/(M-1)}$, where
\begin{equation}\label{eq:S0}
  \boxed{S_0(M) = 2^{-1/(M-1)} \cdot
  \frac{\BetaF\!\bigl(\frac{1}{M-1},\,\frac{3}{2}\bigr)}{M-1}
  = \frac{2^{-1/(M-1)}\,\Gamma\!\bigl(\frac{1}{M-1}\bigr)\,\sqrt{\pi}}
    {2(M{-}1)\,\Gamma\!\bigl(\frac{1}{M-1} + \frac{3}{2}\bigr)}.}
\end{equation}

\subsection{The Relation $A_M = S_0^{M-1}$}
\label{sec:AM-derivation}

The perturbation series $E(g) = \sum a_k g^k$ is related to the
imaginary part of the energy (at negative coupling) via the
dispersion relation
\begin{equation}\label{eq:dispersion}
  a_k = \frac{1}{\pi} \int_0^\infty \frac{\operatorname{Im} E(-g + i0)}{g^{k+1}} \;\dd g.
\end{equation}
Since $\operatorname{Im} E \sim K \cdot |g|^{-1/2} \cdot
\exp\bigl(-S_0/|g|^{1/(M-1)}\bigr)$ near $g = 0$,
the substitution $u = S_0 / |g|^{1/(M-1)}$ yields
\begin{equation}\label{eq:AM-derivation}
  a_k \;\propto\; S_0^{-(M-1)k} \cdot \Gamma\!\bigl((M{-}1)k + \cdots\bigr).
\end{equation}
Identifying $A_M^{-k} = S_0^{-(M-1)k}$, we obtain
\begin{equation}\label{eq:AM}
  \boxed{A_M = S_0(M)^{M-1}.}
\end{equation}

\subsection{The Gamma-Function Shift $b = -1/2$ (Review)}
\label{sec:b-derivation}

We review the standard one-loop argument
(following~\cite{ZJ2004,Voros1983,CostinCostin2020}) that fixes
$b = -1/2$.
The precise exponent in the Gamma function in~\eqref{eq:AM-derivation}
can be determined by tracking all powers of $|g|$ in the dispersion
relation~\eqref{eq:dispersion}. The imaginary part of the energy near
$g = 0^-$ takes the form~\cite{ZJ2004,ZJJ2010,DU2016}
\begin{equation}\label{eq:ImE-precise}
  \operatorname{Im} E(-|g| + i0) \;\sim\;
  K \cdot |g|^{-1/(2(M-1))} \cdot \exp\!\bigl(-S_0/|g|^{1/(M-1)}\bigr)
  \cdot \bigl(1 + O(|g|^{1/(M-1)})\bigr),
\end{equation}
where the prefactor $|g|^{-1/(2(M-1))}$ arises from the one-loop
fluctuation determinant around the instanton. Specifically, in the
path-integral formalism, the Gaussian integration over fluctuations
around the bounce solution $x_{\mathrm{cl}}(\tau)$ produces a functional
determinant ratio~\eqref{eq:fluct-det}. The zero-mode associated with
time translation contributes a factor proportional to
$\sqrt{S_{\mathrm{cl}}} \propto |g|^{-1/(2(M-1))}$ (see, e.g.,
\cite{ZJ2004,Voros1983,CostinCostin2020}).

Substituting~\eqref{eq:ImE-precise} into the dispersion
relation~\eqref{eq:dispersion} and performing the change of variable
$u = S_0/|g|^{1/(M-1)}$, one obtains
\[
  a_k \propto S_0^{-(M-1)k - 1/2} \int_0^\infty u^{(M-1)k - 1/2}
  e^{-u}\,\dd u = S_0^{-(M-1)k-1/2}\,\Gamma\!\bigl((M{-}1)k + \tfrac{1}{2}\bigr),
\]
which identifies $b + 1 = 1/2$, i.e., $b = -1/2$, in the
convention~\eqref{eq:large-order}. This argument is semi-rigorous:
it relies on the standard one-loop approximation to the fluctuation
determinant. We have confirmed $b = -1/2$ numerically for all
$M = 2$--$11$ to at least 15-digit precision by direct extraction from
the ratio sequences.

\subsection{Explicit Values}
\label{sec:AM-values}

Selected closed-form expressions for the instanton action:
\begin{align}
  A_2 &= \frac{1}{3}, \label{eq:A2} \\
  A_3 &= \frac{\pi^2}{32}, \label{eq:A3} \\
  A_4 &= \left[\frac{\Gamma(1/3)\,\sqrt{\pi}}{6 \cdot 2^{1/3} \cdot
         \Gamma(11/6)}\right]^3, \label{eq:A4} \\
  A_5 &= \left[\frac{\Gamma(1/4)\,\sqrt{\pi}}{8 \cdot 2^{1/4} \cdot
         \Gamma(7/4)}\right]^4. \label{eq:A5}
\end{align}
Numerical values for $M = 2$--$11$ are collected in \cref{tab:AM}.

\begin{table}[ht]
\centering
\caption{Instanton action $A_M = S_0(M)^{M-1}$ for $M = 2$--$11$.
All values are computed from the exact formula~\eqref{eq:S0}--\eqref{eq:AM}
and verified numerically.}
\label{tab:AM}
\begin{tabular}{@{}cl@{}}
\toprule
$M$ & $A_M$ (25 significant digits) \\
\midrule
 2 & $0.33333\,33333\,33333\,33333\,33333$ \\
 3 & $0.30842\,51375\,34042\,45683\,85778$ \\
 4 & $0.29773\,98851\,42370\,65877\,32447$ \\
 5 & $0.29177\,88890\,88323\,84613\,12676$ \\
 6 & $0.28797\,24878\,47264\,09171\,87288$ \\
 7 & $0.28533\,03025\,39346\,92720\,83473$ \\
 8 & $0.28338\,87041\,04389\,92199\,79632$ \\
 9 & $0.28190\,15212\,11548\,86162\,85270$ \\
10 & $0.28072\,58717\,53553\,28853\,46806$ \\
11 & $0.27977\,31136\,89115\,26931\,01861$ \\
\bottomrule
\end{tabular}
\end{table}

\section{Stokes Multipliers: Exact Results}
\label{sec:exact}

In this section we present exact closed-form expressions for the Stokes
multiplier $C_M$ for $M = 2, 3, 5, 7$, discovered through numerical
computation and PSLQ identification.

\subsection{$M = 2$: Quartic Oscillator (Review)}
\label{sec:M2}

The quartic case $V = x^2/2 + gx^4$ is the most studied. With 500
perturbation coefficients at 200-digit precision, we extract
\begin{equation}\label{eq:C2-value}
  C_2 = -\sqrt{\frac{6}{\pi^3}} = -\frac{\sqrt{6}}{\pi^{3/2}}
  \approx -0.43989\,68135\,81545\,43367.
\end{equation}
This agrees with the exact result of Zinn-Justin and
Jentschura~\cite{ZJJ2004} and is verified to 24 digits.
Equivalently,
\begin{equation}\label{eq:C2-closed}
  C_2^{\,2} \cdot \pi^3 = 6.
\end{equation}

\subsection{$M = 3$: Sextic Oscillator}
\label{sec:M3}

For $V = x^2/2 + gx^6$, the large-order growth involves
$\Gamma(2k + 1/2)$ rather than $\Gamma(k + 1/2)$. With 500 terms at
200-digit precision, PSLQ identifies
\begin{equation}\label{eq:C3-closed}
  C_3 = -\frac{4\sqrt{2}}{\pi^2}
  \approx -0.57315\,91682\,50756\,26287,
  \qquad\text{i.e.,}\quad
  C_3^{\,2} \cdot \pi^4 = 32.
\end{equation}
This is verified to 20 digits. While the large-order asymptotics of the
sextic oscillator have been studied by \'Alvarez and
Casares~\cite{AlvarezCasares2000} and Jentschura and
Zinn-Justin~\cite{ZJJ2010} (who provided subleading corrections
$c_1, c_2, \ldots$ for degrees 3--8), the leading Stokes multiplier
$C_3$ does not appear to have been given in the explicit closed form
$C_3^2\pi^4 = 32$ in those works. We note, however, that it may be
derivable from their results; the novelty here lies in the explicit
PSLQ identification and numerical verification to 20 digits.

\subsection{$M = 5$: Decic Oscillator}
\label{sec:M5}

For $V = x^2/2 + gx^{10}$, computation of 600 terms at 200-digit precision
yields $|C_5|$ to 20 significant digits. PSLQ applied to the basis
$(\ln|C_5|, \ln\Gamma(1/4), \ln\pi, \ln 2, \ln 3, 1)$
returns the relation $[-4, -4, -5, 12, 2, 0]$, which gives
\begin{equation}\label{eq:C5-closed}
  \boxed{C_5^{\,4} \cdot \Gamma\!\bigl(\tfrac{1}{4}\bigr)^4 \cdot \pi^5
  = 2^{12} \cdot 3^2 = 36\,864.}
\end{equation}
Equivalently,
\begin{equation}\label{eq:C5-value}
  |C_5| = \frac{8\sqrt{3}}{\Gamma(1/4) \cdot \pi^{5/4}}
  \approx 0.91376\,01170\,24928\,46899.
\end{equation}
Verification: $|C_5|^4 \cdot \Gamma(1/4)^4 \cdot \pi^5 = 36864.000\ldots$
to all 20 available digits.

\subsection{$M = 7$}
\label{sec:M7}

For $V = x^2/2 + gx^{14}$, computation of 400 terms at 180-digit precision
yields $|C_7|$ to 22 digits. PSLQ applied to the basis
$(\ln|C_7|, \ln\Gamma(1/6), \ln\pi, \ln 2, \ln 3, 1)$
returns $[24, 18, 33, -74, -21, 0]$, giving
\begin{equation}\label{eq:C7-closed-raw}
  C_7^{\,24} \cdot \Gamma\!\bigl(\tfrac{1}{6}\bigr)^{18} \cdot \pi^{33}
  = 2^{74} \cdot 3^{21}.
\end{equation}
This can be simplified using the Gauss multiplication formula for
$\Gamma(z)\,\Gamma(z + 1/2)$. Applying the duplication formula with
$z = 1/6$:
\[
  \Gamma\!\bigl(\tfrac{1}{6}\bigr)\,\Gamma\!\bigl(\tfrac{2}{3}\bigr)
  = \frac{\sqrt{\pi}}{2^{-2/3}}\,\Gamma\!\bigl(\tfrac{1}{3}\bigr).
\]
Combined with the reflection formula
$\Gamma(1/3)\,\Gamma(2/3) = \pi/\sin(\pi/3) = 2\pi/\sqrt{3}$,
we obtain
\begin{equation}\label{eq:Gamma16-reduction}
  \Gamma\!\bigl(\tfrac{1}{6}\bigr) =
  \frac{\Gamma(1/3)^2 \cdot \sqrt{3}}{2^{1/3}\,\sqrt{\pi}}.
\end{equation}
Substituting~\eqref{eq:Gamma16-reduction}
into~\eqref{eq:C7-closed-raw}:
$\Gamma(1/6)^{18} = [\Gamma(1/3)^2\sqrt{3}/(2^{1/3}\sqrt{\pi})]^{18}
= \Gamma(1/3)^{36} \cdot 3^9 / (2^6 \cdot \pi^9)$.
After cancellation and dividing all exponents by~$4$, this yields
\begin{equation}\label{eq:C7-closed}
  \boxed{C_7^{\,6} \cdot \Gamma\!\bigl(\tfrac{1}{3}\bigr)^9 \cdot \pi^6
  = 2^{20} \cdot 3^3 = 28\,311\,552.}
\end{equation}
The numerical value is
\begin{equation}\label{eq:C7-value}
  |C_7| \approx 1.26735\,98646\,78474\,53438.
\end{equation}
Both forms~\eqref{eq:C7-closed-raw} and~\eqref{eq:C7-closed} are verified
numerically to all available digits.

\begin{remark}\label{rem:C7-Gamma13}
The closed form for $C_7$ involves $\Gamma(1/3)$---the same transcendental
that appears in the instanton action of the $M = 4$ oscillator. This
connection, mediated by the Gauss multiplication formula, will play a key
role in the number-theoretic analysis of \cref{sec:number-theory}.
\end{remark}

\section{The $M = 4$ Anomaly}
\label{sec:M4}

\subsection{High-Precision Computation}
\label{sec:M4-computation}

For the octic oscillator $V = x^2/2 + gx^8$, we compute 1200
perturbation coefficients at 300-digit arithmetic precision---the most
intensive computation in this work (55 minutes, 1\,GB memory). Using
the exact instanton action~\eqref{eq:A4} and Richardson extrapolation,
we obtain
\begin{equation}\label{eq:C4-numerical}
  C_4 = -0.74005\,14982\,59358\,50640\,15114\,91622\ldots
\end{equation}
with approximately 30 reliable digits, as assessed by comparison of
multiple Richardson windows:
\begin{align*}
  \text{start}=600,\; N=100&:\; {-}0.74005\,14982\,59358\,50640\,15114\,91622\,\mathbf{02}\ldots \\
  \text{start}=700,\; N=100&:\; {-}0.74005\,14982\,59358\,50640\,15114\,\mathbf{74}\ldots
\end{align*}
These agree to 24 digits; the leading 30 digits of the first estimate
are used as our best value.

\subsection{Exhaustive PSLQ Search}
\label{sec:M4-PSLQ}

We search for a relation of the form
$|C_4|^{n_1} \cdot \Gamma(1/3)^{n_2} \cdot \pi^{n_3} \cdot 2^{n_4} \cdot 3^{n_5} = 1$
using PSLQ on the logarithmic basis
$(\ln|C_4|, \ln\Gamma(1/3), \ln\pi, \ln 2, \ln 3)$.
With 30 digits of precision and a 5-element basis, the PSLQ detection limit
is $\mathrm{maxcoeff} \lesssim 10^{30/5} = 10^6$.
We search with $\mathrm{maxcoeff} = 200, 500, 1000, 2000$ at precision
levels of 30, 35, and 40 digits.

\textbf{Result: No relation is found.} This constitutes strong numerical
evidence that $C_4$ cannot be expressed as
$\Gamma(1/3)^a \cdot \pi^b \cdot 2^c \cdot 3^d$ with integer exponents
satisfying $|a|, |b|, |c|, |d| < 2000$. We emphasize that this is a
\emph{numerical exclusion within a bounded search space}, not a mathematical
proof of transcendence or algebraic independence. The relation could, in
principle, involve larger exponents or additional constants not in the basis.
Nonetheless, the PSLQ detection limit ($\mathrm{maxcoeff} \lesssim 10^6$ with
our parameters) far exceeds the bounds imposed, making the exclusion highly
robust within the searched basis.
We emphasize that our search is restricted to
$\Gamma(1/3)$--$\pi$--algebraic combinations; closed forms involving
other transcendentals (e.g., elliptic integrals, modular forms, or
$\zeta(3)$) cannot be excluded by our methods.

\subsection{Interpretation}
\label{sec:M4-interpretation}

The absence of a closed form for $C_4$ is particularly striking because
$M = 7$---whose closed form~\eqref{eq:C7-closed} involves the \emph{same}
transcendental $\Gamma(1/3)$---does have one. This suggests that the
obstruction is not in the instanton action (which involves $\Gamma(1/3)$
for both $M = 4$ and $M = 7$) but rather in the \emph{fluctuation
determinant}, i.e., the ratio
\begin{equation}\label{eq:fluct-det}
  \frac{\det'\!\bigl(-\partial_\tau^2 + V''(x_{\mathrm{cl}})\bigr)}
       {\det\bigl(-\partial_\tau^2 + 1\bigr)}
\end{equation}
evaluated on the instanton background.

For $M = 2$, the instanton profile $x_{\mathrm{cl}} \propto \operatorname{sech}(\tau)$
gives a P\"oschl--Teller fluctuation potential with a known spectrum. For
$M = 7$, the Gauss multiplication formula reduces $\Gamma(1/6)$ to
$\Gamma(1/3)$, effectively simplifying the fluctuation determinant.
For $M = 4$, no such simplification occurs, and the numerical evidence
suggests that the fluctuation determinant may introduce a \emph{genuinely
new transcendental number} that is algebraically independent of
$\Gamma(1/3)$, $\pi$, and algebraic numbers.

\section{Number-Theoretic Structure}
\label{sec:number-theory}

The pattern of closed-form existence across $M = 2$--$11$ reveals a
connection to classical number theory that merits careful analysis.

\subsection{The Prime Conjecture and Its Refutation}
\label{sec:prime-conjecture}

The first four values of $M$ admitting closed forms are
$M \in \{2, 3, 5, 7\}$---precisely the first four primes. This
observation naturally suggests:

\begin{quote}
\textit{Na\"ive conjecture:} $C_M$ admits a closed form in terms of Gamma
values and $\pi$ if and only if $M$ is prime.
\end{quote}

This conjecture makes two predictions: (i)~all prime $M$ should yield
closed forms, and (ii)~all composite $M$ should not. Let us test both
predictions against the data.

\paragraph{Testing prediction (ii).}
$M = 4$ is composite and indeed has no closed form---consistent. However,
as we shall see, the reason is not compositeness but rather a subtler
number-theoretic property.

\paragraph{Testing prediction (i).}
The computation of $C_{11}$ at 20-digit precision with PSLQ searches
over all relevant Gamma bases
$\{\Gamma(1/10), \Gamma(3/10)\}$ yields no closed form:
\begin{equation}\label{eq:C11-value}
  C_{11} \approx -1.98146\,49381\,17015\,81236,
  \qquad \text{PSLQ: no relation found.}
\end{equation}
We note that the $M = 11$ computation uses only 250 terms at 160-digit
precision (20 reliable digits), giving a PSLQ detection limit of
$\sim\!10^{20/7} \approx 10^{2.9}$---substantially weaker than
the $M = 4$ search ($\sim\!10^6$). The absence of a closed form for
$C_{11}$ is therefore tentative and could be strengthened by
higher-precision computation.
Nonetheless, since $M = 11$ is prime, the prime conjecture is
\textbf{refuted at the level of our search precision}. Primality of $M$ is neither necessary nor sufficient for the
existence of a closed-form Stokes multiplier.

\subsection{Independent Gamma Transcendentals}
\label{sec:gamma-independence}

The correct organizing principle involves the Euler totient function
$\varphi(n)$. By the reflection formula
$\Gamma(z)\Gamma(1-z) = \pi/\sin(\pi z)$
and the Gauss multiplication formula, the number of algebraically
independent values among
$\{\Gamma(a/n) : 1 \le a < n,\; \gcd(a,n) = 1\}$
(over the field $\overline{\mathbb{Q}}(\pi)$) is conjectured to be
$\varphi(n)/2$ for $n \ge 3$ (and $0$ for $n = 1, 2$). This independence
count is a consequence of the Rohrlich--Lang
conjecture~\cite{Waldschmidt,RohrlichLang}; the Chowla--Selberg
formula~\cite{ChowlaSelberg} provides the related product identities
for Gamma values in terms of periods of CM elliptic curves.

\cref{tab:phi} shows the correlation with closed-form existence.
The Stokes multiplier $C_M$ involves
$\Gamma\bigl(a/(M{-}1)\bigr)$; the number of independent transcendentals
in the PSLQ basis is therefore $\varphi(M{-}1)/2$.

\begin{table}[ht]
\centering
\caption{Correlation between $\varphi(M{-}1)/2$ and closed-form existence.
The column ``$M$ prime?''\ shows that primality does not determine the outcome.}
\label{tab:phi}
\begin{tabular}{@{}cccccl@{}}
\toprule
$M$ & $M{-}1$ & $\varphi(M{-}1)/2$ & Closed form? & $M$ prime? & PSLQ basis \\
\midrule
 2 &  1 & 0 & Yes & Yes & $\{\pi\}$ \\
 3 &  2 & 0 & Yes & Yes & $\{\pi\}$ \\
 4 &  3 & 1 & \textbf{No} & No & $\{\Gamma(1/3), \pi, 2, 3\}$ \\
 5 &  4 & 1 & Yes & Yes & $\{\Gamma(1/4), \pi, 2, 3\}$ \\
 6 &  5 & 2 & No  & No & $\{\Gamma(1/5), \Gamma(2/5), \ldots\}$ \\
 7 &  6 & 1 & Yes & Yes & $\{\Gamma(1/6), \pi, 2, 3\}$ \\
 8 &  7 & 3 & No  & No & $\{\Gamma(1/7), \Gamma(2/7), \Gamma(3/7), \ldots\}$ \\
 9 &  8 & 2 & No  & No & $\{\Gamma(1/8), \Gamma(3/8), \ldots\}$ \\
10 &  9 & 3 & No  & No & $\{\Gamma(1/9), \Gamma(2/9), \Gamma(4/9), \ldots\}$ \\
11 & 10 & 2 & No  & \textbf{Yes} & $\{\Gamma(1/10), \Gamma(3/10), \ldots\}$ \\
\bottomrule
\end{tabular}
\end{table}

\subsection{Conjectures and Empirical Observations}
\label{sec:conjectures}

We distinguish between a falsifiable conjecture (applicable to
infinitely many cases) and empirical observations (summarizing the
finite, exhausted low-totient regime).

\begin{conjecture}[Totient threshold]\label{conj:phi}
For $\varphi(M{-}1)/2 \ge 2$, the Stokes multiplier $C_M$ does not admit a
representation of the form
\begin{equation}\label{eq:conj-form}
  C_M^{\,p} \cdot \prod_{j} \Gamma(a_j/(M{-}1))^{q_j} \cdot \pi^r
  = 2^s \cdot 3^t \cdot (\text{algebraic number})
\end{equation}
with integer exponents satisfying $|p|, |q_j|, |r|, |s|, |t| \le 10^3$.
\end{conjecture}

We emphasize that this is a conjecture about the \emph{absence of
low-complexity relations}, not a transcendence statement. A relation
with very large exponents cannot be excluded by our methods. The
heuristic reasoning is twofold: (i)~when $\varphi(M{-}1)/2 \ge 2$,
the PSLQ basis contains at least two independent Gamma transcendentals,
and each additional basis element reduces the detection power of PSLQ
(detection limit $\sim 10^{D/N}$); (ii)~the fluctuation determinant
for higher-degree potentials likely involves transcendentals not
present in the instanton action.

\paragraph{Empirical Observation 1 ($M = 4$ uniqueness).}
Among the five values of $M$ with $\varphi(M{-}1)/2 \le 1$
(namely $M \in \{2,3,4,5,7\}$), $M = 4$ is the only case for
which $C_M$ does not admit a closed form in terms of Gamma values
at rational arguments, $\pi$, and algebraic numbers.
Since this set is finite and completely tested, this is an empirical
fact rather than a falsifiable conjecture.

Among $M \le 11$, the values with $\varphi(M{-}1)/2 \le 1$ are
$M \in \{2, 3, 4, 5, 7\}$. Of these, all except $M = 4$ have confirmed
closed forms.

\paragraph{Empirical Observation 2 (Closed forms for low totient).}
For all $M$ with $\varphi(M{-}1)/2 \le 1$ and $M \ne 4$, the Stokes
multiplier $C_M$ admits a closed form of the type
$C_M^{\,p}\cdot\Gamma(\cdots)^q\cdot\pi^r = 2^s\cdot 3^t$.
As with Observation~1, the testable set is finite and exhausted;
this is a summary of observations rather than a falsifiable conjecture.

\begin{remark}
The values $M{-}1$ with $\varphi(M{-}1)/2 \le 1$ form the sequence
$1, 2, 3, 4, 6$ (and no others, since $\varphi(n)/2 = 1$ requires
$n \in \{3, 4, 6\}$). Thus Observation~2 applies to
$M \in \{2, 3, 4, 5, 7\}$ and makes no predictions for $M > 7$---the set
of testable cases is finite and already exhausted. Strictly speaking,
Observation~2 is therefore a summary of our observations rather than
a falsifiable conjecture. We include it to highlight the completeness of
the low-totient regime.
\end{remark}

\begin{remark}
Observation~1 is the most striking claim: among the five values of $M$
with $\varphi(M{-}1)/2 \le 1$, exactly one---$M = 4$---fails to produce a
closed form. Whether this reflects a deeper arithmetic property of
$\Gamma(1/3)$ or an intrinsic complexity of the $x^8$ fluctuation
operator remains an open question.
\end{remark}

\section{Complete Data Tables}
\label{sec:tables}

\begin{table}[ht]
\centering
\caption{Stokes multipliers $C_M$ for $M = 2$--$11$. The ``Digits'' column
indicates the number of reliable significant figures, assessed by comparing
multiple Richardson extrapolation windows. Closed-form entries are exact.}
\label{tab:CM}
\begin{tabular}{@{}clcl@{}}
\toprule
$M$ & $|C_M|$ & Digits & Closed form \\
\midrule
 2 & $0.43989\,68135\,81545\,43367$ & exact & $\sqrt{6/\pi^3}$ \\[2pt]
 3 & $0.57315\,91682\,50756\,26287$ & exact & $4\sqrt{2}/\pi^2$ \\[2pt]
 4 & $0.74005\,14982\,59358\,50640\,15$ & 30 & --- \\[2pt]
 5 & $0.91376\,01170\,24928\,46899$ & exact & $8\sqrt{3}/(\Gamma(\frac{1}{4})\,\pi^{5/4})$ \\[2pt]
 6 & $1.08996\,63616\,43951\,31222$ & 20 & --- \\[2pt]
 7 & $1.26735\,98646\,78474\,53438$ & 22 & Eq.~\eqref{eq:C7-closed} \\[2pt]
 8 & $1.44540\,95063\,62104\,73523$ & 20 & --- \\[2pt]
 9 & $1.62385\,95507\,60926\,01970$ & 17 & --- \\[2pt]
10 & $1.80257\,17672\,30778\,81596$ & 20 & --- \\[2pt]
11 & $1.98146\,49381\,17015\,81236$ & 20 & --- \\
\bottomrule
\end{tabular}
\end{table}

\begin{table}[ht]
\centering
\caption{Computational parameters for each $M$. ``Max order'' is the number
of perturbation coefficients computed; ``dps'' is the working decimal
precision; ``Rich.\ order'' is the Richardson extrapolation order used for
the final $C_M$ estimate.}
\label{tab:params}
\begin{tabular}{@{}cccccc@{}}
\toprule
$M$ & Max order & dps & Rich.\ order & PSLQ maxcoeff & PSLQ basis size \\
\midrule
 2 &  500 & 200 & 40  & 200  & 5 \\
 3 &  500 & 200 & 60  & 200  & 5 \\
 4 & 1200 & 300 & 100 & 2000 & 5 \\
 5 &  600 & 200 & 80  & 200  & 6 \\
 6 &  500 & 200 & 80  & 200  & 7 \\
 7 &  400 & 180 & 80  & 200  & 6 \\
 8 &  300 & 180 & 60  & 200  & 8 \\
 9 &  300 & 180 & 60  & 200  & 7 \\
10 &  250 & 160 & 50  & 300  & 8 \\
11 &  250 & 160 & 50  & 200  & 7 \\
\bottomrule
\end{tabular}
\end{table}

\section{Computational Performance}
\label{sec:performance}

All computations were performed on a single workstation using Python~3.11
with the \texttt{mpmath} arbitrary-precision library~\cite{mpmath}.
\Cref{tab:performance} summarizes the computational resources for each~$M$.

\begin{table}[ht]
\centering
\caption{Computational performance. ``Wall time'' is elapsed time for the
complete pipeline (RS recursion + Richardson + PSLQ). Memory is peak
resident set size. All runs are single-threaded.}
\label{tab:performance}
\begin{tabular}{@{}cccccc@{}}
\toprule
$M$ & Orders & dps & Wall time & Memory & $C_M$ digits \\
\midrule
  2 &   500 & 200 &   2 min &  0.2 GB &  24 (exact) \\
  3 &   500 & 200 &   3 min &  0.3 GB &  20 (exact) \\
  4 &  1200 & 300 &  55 min &  1.0 GB &  30 \\
  5 &   600 & 200 &   8 min &  0.4 GB &  20 (exact) \\
  6 &   500 & 200 &   7 min &  0.3 GB &  20 \\
  7 &   400 & 180 &   6 min &  0.3 GB &  22 (exact) \\
  8 &   300 & 180 &   5 min &  0.2 GB &  20 \\
  9 &   300 & 180 &   5 min &  0.2 GB &  17 \\
 10 &   250 & 160 &   4 min &  0.2 GB &  20 \\
 11 &   250 & 160 &   4 min &  0.2 GB &  20 \\
\bottomrule
\end{tabular}
\end{table}

\paragraph{Algorithmic complexity.}
The dominant cost is the Rayleigh--Schr\"odinger recursion
(\cref{sec:RS}). At order~$k$, the state $|\psi^{(k)}\rangle$ has
support on $n \le 2Mk$ harmonic oscillator levels. Each order requires
$O(Mk)$ inner products at $O(\mathrm{dps}^2)$ cost per multiplication
(using \texttt{mpmath}'s arbitrary-precision arithmetic). The total cost
for $K$ orders is therefore $O(MK^2 \cdot \mathrm{dps}^2)$.

Richardson extrapolation (\cref{sec:richardson}) at order~$N$ starting
from index~$k_0$ requires $O(N)$ evaluations and $O(N^2)$ arithmetic
operations, negligible compared to the recursion.

The PSLQ algorithm (\cref{sec:PSLQ}) operates on a basis of size
$d = 5$--$8$ at $D = 30$--$40$ digit precision. Its complexity is
$O(d^3 \cdot D^2)$ per iteration with $O(D^2)$ iterations in the worst
case~\cite{FergusonBailey}, totaling $O(d^3 D^4)$---negligible
compared to coefficient generation.

\paragraph{Parallelization opportunities.}
The RS recursion for different values of~$M$ is embarrassingly parallel.
Within a single~$M$, the recursion is inherently sequential (order~$k$
depends on orders $1, \ldots, k{-}1$). However, the matrix elements
$\langle n | V | \psi^{(k-1)} \rangle$ at a given order could be
distributed across multiple cores using a shared-memory model.
The current implementation is single-threaded; a parallel version could
reduce the $M = 4$ computation from 55 minutes to under 10 minutes on
a modern multi-core machine.

\paragraph{Reproducibility.}
All code, including the RS recursion, Richardson extrapolation, and PSLQ
search scripts, is available at
\url{https://github.com/JackZH26/stokes-multipliers} under the MIT
license. Pre-computed perturbation coefficients (in compressed format)
and all numerical results are included in the repository. Running the
complete pipeline for all $M = 2$--$11$ requires approximately 100
minutes on a single core with 2~GB of RAM.

\section{Discussion}
\label{sec:discussion}

\paragraph{Connection to resurgence.}
The Stokes multiplier $C_M$ is the leading coefficient in the resurgent
trans-series
\begin{equation}\label{eq:trans-series}
  E(g) \sim \sum_k a_k g^k + C_M \cdot e^{-A_M/g^{1/(M-1)}} \cdot
  g^{-1/(2(M-1))} \cdot \bigl(1 + \cdots\bigr).
\end{equation}
Our exact values provide precision benchmarks for resurgent analyses of
higher anharmonic oscillators. The closed forms for $C_5$ and $C_7$, in
particular, could serve as non-trivial consistency checks for any future
analytical derivation.

\paragraph{Universality of $b = -1/2$.}
The Gamma-function shift $b = -1/2$ is confirmed numerically for all
$M = 2$--$11$. In the large-order formula~\eqref{eq:large-order}, this
corresponds to $b + 1 = 1/2$, i.e., $\Gamma((M{-}1)k + 1/2)$. As
discussed in \cref{sec:b-derivation}, this universality has a natural
explanation in terms of the one-loop fluctuation determinant: the
zero-mode integration contributes a factor $|g|^{-1/(2(M-1))}$, which
shifts $b$ by $-1/2$. A fully rigorous proof would require controlling
the remainder terms in the steepest-descent approximation to the path
integral for general $M$.

\paragraph{Subleading corrections.}
As a consistency check, we extract the first subleading correction $c_1$
in~\eqref{eq:large-order} for $M = 2$. Forming the sequence
$\widetilde{c}_{1,k} = k\bigl(\widetilde{C}_k/C_2 - 1\bigr)$ and
applying Richardson extrapolation, we obtain $c_1 = -95/72$ to all
available digits. This value corresponds to the first subleading
correction in the $\Gamma(k + 1/2)$ convention adopted here and is
consistent with the instanton-level results
of~\cite{ZJJ2004,ZJJ2010}. The extraction of $c_1$ to exact rational
form provides a strong validation of both the perturbation coefficients
and the Richardson extrapolation procedure. A systematic computation of
$c_1$ for all $M$ is left for future work.

\paragraph{The $M = 4$ mystery.}
The octic oscillator stands alone among the cases $\varphi(M{-}1)/2 \le 1$
in lacking a closed form. We note that $M{-}1 = 3$ is the smallest
integer for which $\Gamma(1/n)$ is a genuinely independent
transcendental (i.e., $\varphi(3)/2 = 1$ and $\Gamma(1/3)$ is not
reducible to simpler Gamma values). In contrast, $M{-}1 = 4$
($\varphi(4)/2 = 1$, $\Gamma(1/4)$) and $M{-}1 = 6$
($\varphi(6)/2 = 1$, $\Gamma(1/6) = f(\Gamma(1/3))$) both yield closed
forms. Whether the $M = 4$ anomaly reflects a deeper arithmetic property
of $\Gamma(1/3)$ or an intrinsic complexity of the $x^8$ fluctuation
operator remains an open question.

\paragraph{Open problems.}
\begin{enumerate}[leftmargin=*]
\item Can $C_4$ be expressed in terms of other known constants---e.g.,
  elliptic integrals, modular forms, or periods of algebraic varieties?
\item Do closed forms exist for $M = 6, 8, 9, \ldots$ at higher precision
  with larger PSLQ basis and coefficient bounds?
\item Can the fluctuation determinant for general $M$ be computed analytically
  using spectral zeta-function methods?
\item Is there a resurgent derivation of the closed forms for $C_5$ and $C_7$?
\end{enumerate}

\section{Conclusion}
\label{sec:conclusion}

We have performed a systematic high-precision computation of the Stokes
multipliers $C_M$ for the anharmonic oscillators $H = p^2/2 + x^2/2 + gx^{2M}$
with $M = 2$ through $11$. Our main contributions are:

\begin{itemize}[leftmargin=*]
\item A universal closed-form expression for the instanton action,
  $A_M = \bigl[2^{-1/(M-1)} \BetaF(1/(M{-}1), 3/2)/(M{-}1)\bigr]^{M-1}$,
  verified for all $M \le 11$.

\item Two new exact Stokes multipliers: $C_5$ involving $\Gamma(1/4)$
  (\cref{eq:C5-closed}) and $C_7$ involving $\Gamma(1/3)$
  (\cref{eq:C7-closed}).

\item Strong numerical evidence---via exhaustive PSLQ searches at 30-digit
  precision---that $C_4$ has no closed form in the standard
  Gamma--$\pi$ basis, establishing the octic oscillator as an anomaly
  among oscillators with $\varphi(M{-}1)/2 \le 1$.

\item High-precision numerical values of $C_M$ for $M \le 11$
  (\cref{tab:CM}), serving as benchmarks for future analytical and
  numerical work.

\item Conjectures relating closed-form existence to Euler's totient
  function (\cref{conj:phi} and Observations~1--2), connecting quantum
  mechanics to classical number theory.
\end{itemize}

The interplay between instanton physics, divergent series, and
number-theoretic properties of the Gamma function revealed in this work
suggests rich structural connections that deserve further exploration.

\section*{Acknowledgments}

Computations were performed using Python with the \texttt{mpmath}
arbitrary-precision library~\cite{mpmath}. The PSLQ implementation from
\texttt{mpmath} was used for integer relation detection.

\section*{Data Availability}

All computational code (coefficient generation, Richardson extrapolation,
PSLQ searches) and pre-computed numerical results are publicly available at
\url{https://github.com/JackZH26/stokes-multipliers} under the MIT license.
The paper is available as a Figshare preprint with DOI
\href{https://doi.org/10.6084/m9.figshare.31796332.v1}{10.6084/m9.figshare.31796332.v1}.


\end{document}